\documentclass{emulateapj}
\usepackage{apjfonts}

\input psfig

\newcommand{\etal}{et~al.~}
\def\ts{\thinspace}
 
\def\lapprox{$_<\atop{^\sim}$}
\def\omit#1{\empty}

\newdimen\sa  \def\sd{\sa=.1em    \ifmmode $\rlap{.}$''$\kern -\sa$
                                      \else \rlap{.}$''$\kern -\sa\fi}
\newdimen\sb  \def\md{\sb=0.07em  \ifmmode $\rlap{.}$'$\kern -\sb$
                                      \else \rlap{.}$'$\kern -\sb\fi}

\begin{document}

\lefthead{Pseudobulges in NGC 7690 and NGC 4593}

\righthead{Kormendy et al.}

\title{Pseudobulges in the Disk Galaxies NGC 7690 and
       NGC 4593\altaffilmark{1,2}}

\altaffiltext{1}{Based on observations made with the 
       Anglo-Australian Telescope.}
\altaffiltext{2}{Based in part on 
       observations made with the NASA/ESA Hubble Space Telescope, 
       obtained from the Data Archive at the Space Telescope Science  
       Institute (STScI).  STScI is operated by the Association of 
       Universities for Research in Astronomy, Inc., under NASA 
       contract NAS5--26555.  The observations of NGC 7690 are 
       associated with program IDs 7331 (NICMOS -- Massimo Stiavelli)
       and 6359 (WFPC2 -- Massimo Stiavelli).  The observations of
       NGC 4593 are associated with program IDs 7330 (NICMOS -- John 
       Mulchaey), and 5479 (WFPC2 -- Matthew Malkan).
       }

\author{  
John Kormendy\altaffilmark{3,4}, 
Mark E.~Cornell\altaffilmark{3}, 
David L.~Block\altaffilmark{4}, 
Johan H.~Knapen\altaffilmark{5},
Emma L.~Allard\altaffilmark{5}
}

\altaffiltext{3}{Department of Astronomy, University of Texas, 1 University Station,
Austin, Texas 78712-0259; kormendy@astro.as.utexas.edu, cornell@astro.as.utexas.edu
}

\altaffiltext{4}{Cosmic Dust Laboratory, School of Computational and
Applied Mathematics, University of the Witwatersrand, Private Bag 3, WITS
2050, Johannesburg, South Africa; block@cam.wits.ac.za}

\altaffiltext{5}{Centre for Astrophysics Research, University of Hertfordshire,
Hatfield, Herts AL10 9AB, United Kingdom; j.knapen@star.herts.ac.uk,
allard@star.herts.ac.uk
}

\pretolerance=15000  \tolerance=15000

\begin{abstract} 
      We present $K_{\rm s}$-band surface photometry of NGC 7690 (Hubble type 
Sab) and NGC 4593 (SBb).  We find that, in both galaxies, a major part of
the ``bulge'' is as flat as the disk and has approximately the same color
as the inner disk.  In other words, the ``bulges'' of these galaxies have 
disk-like properties. We conclude that these are examples of ``pseudobulges''
-- that is, products of secular dynamical evolution.  Nonaxisymmetries such 
as bars and oval disks transport disk gas toward the center.  There, star 
formation builds dense stellar components that look like -- and often are 
mistaken for -- merger-built bulges but that were constructed slowly out of 
disk material. These pseudobulges can most easily be recognized when, 
as in the present galaxies, they retain disk-like properties.  NGC 7690 
and NGC 4593 therefore contribute to the growing evidence that secular 
processes help to shape galaxies.

      NGC 4593 contains a nuclear ring of dust that is morphologically 
similar to nuclear rings of star formation that are seen in many barred and 
oval galaxies.  The nuclear dust ring is connected to nearly radial dust 
lanes in the galaxy's bar.  Such dust lanes are a signature of gas inflow.
We suggest that gas is currently accumulating in the dust ring and 
hypothesize that the gas ring will starburst in the future. The observations 
of NGC 4593 therefore suggest that major starburst events that contribute
to pseudobulge growth can be episodic.
\end{abstract}


\section{Introduction}

\pretolerance=15000 \tolerance=15000

    Internal secular evolution of galaxies is the dynamical redistribution 
of energy and angular momentum that causes galaxies to evolve slowly 
between rapid (collapse-timescale) transformation events that are
caused by galaxy mergers.  Driving agents include nonaxisymmetries in
the gravitational potential such as bars, oval disks, and global spiral
structure.  Kormendy (1993) and Kormendy \& Kennicutt (2004, hereafter KK)
review the growing evidence that secular processes have shaped the 
structure of many galaxies. 

      The fundamental way that self-gravitating disks evolve~-- provided
that there is an efficient driving agent -- is by spreading (Lynden-Bell
\& Kalnajs 1972; Lynden-Bell \& Pringle 1974; Tremaine 1989; see Kormendy
\& Fisher 2005 for a review in the present context).  In general, it is
energetically favorable to shrink the inner parts by expanding the outer
parts.  In barred galaxies, one well known consequence is the production 
of ``inner rings'' around the end of the bar and ``outer rings'' at about 
2.2 bar radii.  The most general consequence of secular evolution, and the
one that is of interest in this paper, is that some disk gas is driven to 
small radii where it reaches high densities, feeds starbursts, and builds
``pseudobulges''.  Because of their high stellar densities and steep
density gradients, pseudobulges superficially resemble -- and often are 
mistaken for -- bulges.  Following Sandage (1961) and Sandage \& Bedke 
(1994), Renzini (1999) adopts this definition of a bulge: ``It appears 
legitimate to look at bulges as ellipticals that happen to have a 
prominent disk around them [and] ellipticals as bulges that for some 
reason have missed the opportunity to acquire or maintain a prominent 
disk.''  We adopt the same definition.  Our paradigm of galaxy formation 
then is that bulges and ellipticals 
both formed via galaxy mergers (e.{\ts}g., Toomre 1977; Steinmetz \& 
Navarro 2002, 2003), a conclusion that is well supported by observations 
(see Schweizer 1990 for a review).  Pseudobulges are therefore 
fundamentally different from bulges -- they were built slowly out of the 
disk. Two well developed examples, one in the unbarred galaxy NGC 7690 
and one in the barred galaxy NGC 4593, are the subjects
of this paper.

    Hierarchical clustering and galaxy merging are well known.
Secular evolution is less studied and less well known. We are 
therefore still in the ``proof of concept'' phase in which it is useful
to illustrate clearcut examples of the results of
secular evolution.  This paper continues a series (see the above 
reviews and Kormendy \& Cornell 2004) in which we illustrate the
variety of disk-like features that define pseudobulges.


\section{Observations and Data Reductions}

\subsection{AAT Infrared Imaging and Data Reduction}

      NGC 7690 and NGC 4593 show dust absorption features near their 
centers.  Also, if star formation were in progress, there would be a
danger that the brightness distributions at visible wavelengths would
be affected by strong variations in mass-to-light ratios.  We therefore
base our results on near-infrared photometry, and we check later that 
they are not greatly affected by stellar population gradients or by dust 
absorption.

      Near-infrared images of both galaxies were obtained with
the InfraRed Imaging Spectrograph 2 (IRIS2; Tinney et al.~2004) at the
$f/8$ Cassegrain focus of the Anglo-Australian Telescope (AAT). We used
IRIS2 in its wide-field imaging mode; this provides a field of view
of 7\md7 $\times$ 7\md7 sampled at a scale of 0\sd4486 pixel$^{-1}$.
The bandpass was $K_{\rm s}$ for both galaxies.  We imaged NGC~7690 for
a total on-source exposure time of 56 m.  These observations were taken 
on 2004 July 1.  For NGC 4593, we obtained 52 m of on-source exposure
piecewise during the nights of 2004 June 30, July 1, July 2, and July 4.
Individual images from different nights were combined during the
reduction process.

       Our observing techniques were similar to those described 
in Knapen et al.~(2003) and in Block et al.~(2004). Individual 8~s
exposures were co-added into raw images of 56 s exposure time. 
In contrast to our earlier work and taking advantage 
of the large IRIS2 field of view compared to the size of the galaxies, 
we did not alternate telescope pointings between the target and a 
nearby, blank background field.  Instead, we used a grid of four 
pointings, each of which imaged the galaxy in one quadrant of the 
detector array.  This ``quadrant-jitter'' method has the advantage of 
high observing efficiency, because no separate background exposures 
are needed, but the disadvantage that the galaxy image must be 
removed from the raw exposures when constructing background sky images. 

\omit{
 We used two independent methods for the
background subtraction. The first is based on the procedures described
in more detail by Knapen et al. (2003), where we used an iterative
median averaging of a small number of frames to remove the galaxy
image and produce a sky background frame for each 5-10~minute block of
observations, depending on the fluctuations with time of the
background. This process resulted in the appearance of ghost images in
the sky background at the locations of the original galaxy images (one
in each of the quadrants, which thus led to nine identical ghosts in
the final, combined, image). These ghosts, at a level of at most 0.3\%
of the sky background, were then isolated, compared, combined, and
finally subtracted from the final image. After flat fielding, sky
subtraction, and combining, all following Knapen et al. (2003),
and ghost-correcting, the resulting images have a useful size of
some 3.5~arcmin. The spatial resolution, as measured from Gaussian
fits to foreground stars, is 1.6, 1.0, and 1.3~arcsec, respectively,
in the $J, H$ and $K_{\rm s}$ images of NGC~7690, and 1.3~arcsec in
the $K_{\rm s}$ image of NGC~4593.  In the second method, we
constructed sky background images by masking out the area of the
galaxy in each of the individual one minute exposures, before
combining the masked frames. We used this method to confirm the
results of the first method, and found that the discrepancies between
the resulting images, visible only in radial profiles at large
galactocentric radii, were smaller than the errors in the estimation
of the final background level.  ***JHK to add photometric calibration
and description of noise level  in images, plus compare to 2MASS and
my own earlier data*** 
}

      The AAT $K_{\rm s}$-band images were reduced in {\tt IRAF\/} (Tody 
1986) using our own procedures. The NGC 7690 data reductions were relatively
simple, because the images were obtained at low airmass on a single 
photometric night.  The galaxy was observed in quadrant-jitter mode, so 
it was centered in a different quadrant of the detector in each frame of a 
consecutive series.  We constructed a flat-field frame for each 
detector quadrant from the median of all images with no galaxy in that
quadrant.  The images were adjusted for any additive offset before computing 
the median.  The NGC 7690 field had few stars, and none remained in the final
flat field.  Once the quadrants were flattened, the galaxy quadrants were 
registered and stacked using a median combine, again correcting for additive 
offsets.  The final stacked frame was quite flat, so a single constant sky 
value was subtracted.  In contrast, the NGC 4593 images were taken over 4 
nights, at high airmass, and in more variable conditions.  The background 
shape varied enough so that the simple assumptions used for NGC 7690 did 
not work for NGC 4593.  We constructed a flat-field frame from all of the 
NGC 4593 images as before.  After flat-fielding, we subtracted
from each galaxy frame a sky frame constructed from the median of the 
4 frames taken nearest in time to the galaxy frame but with the galaxy
in a different quadrant.  Stars in the field were masked before constructing
the flat-field and sky frames. 

     The reduced NGC 7690 image has
PSF \hbox{FWHM = 1\farcs0} and the NGC 4593 image has FWHM = 1\farcs6.

\subsection{Additional Archival Images}

      The images used for surface photometry were archival {\it Hubble Space 
Telescope\/} ({\it HST\/}) WFPC2 F606W images, {\it HST\/} NICMOS F160W 
images, archival 2MASS images (extended 
source catalog tile for NGC 7690 and a Large Galaxy Atlas image for NGC
4593), and the AAT  $K_{\rm s}$-band images obtained with IRIS2. 
To improve signal-to-noise, the 2MASS $J$, $H$, and 
$K_{\rm s}$ images were added together.  The background was removed from
the NGC 7690 2MASS tile by masking the stars and then fitting and 
subtracting a quadratic surface.  The background in the 2MASS image of
NGC 4593 was already sufficiently flat. 

Cosmic ray hits and bad pixels in the NICMOS data were removed using the 
{\tt tvzap} command in Jon Holtzman's
({\tt http://astronomy.nmsu.edu/holtz/xvista\/}) implementation of 
{\tt VISTA\/} (Lauer et al.~1983; Stover 1988).  This 
replaces user-selected pixels with the median of the surrounding 
5 $\times$ 5 pixels.  Cosmic ray hits in the WFPC2 image of NGC 7690  
were removed using the {\tt STSDAS} task {\tt CRREJ}.  The central, 
bad 1 -- 2 columns in the NICMOS data were fixed by linearly interpolating
the neighboring pixel values line by line.  Gaps in the mosaiced
Wide Field Camera image were also filled via linear interpolation.
The {\it HST\/} PC $V$-band image of NGC 4593 that is used in Figure 3 
was cleaned of cosmic rays using the {\tt IRAF\/} script {\tt L.A.COSMIC\/} 
(van Dokkum 2001) and {\tt VISTA tvzap\/}.

Finally, any remaining sky background level was computed as the average 
of the modes of the pixel values in several sky boxes chosen to be free
of galaxy light or interfering objects.  The galaxies fill
the {\it HST\/} Planetary Camera (PC) and NICMOS fields of view.  
For the PC, the sky flux was measured on the Wide Field Camera images and 
scaled to the PC pixels.  For the NICMOS image, the sky was taken as zero 
for NGC 7690.  For NGC 4593, the sky value for the NICMOS image was chosen 
to optimize the agreement between the major-axis cuts as measured on the
NICMOS and AAT images in the radius range $3^{\prime\prime} \leq r \leq
13^{\prime\prime}$.

\subsection{Surface Photometry}

Before fitting ellipses and calculating profile cuts, interfering 
foreground and background objects were identified using Source
Extractor (Bertin \& Arnouts 1996) and masked. Any remaining
stars were identified visually and masked as well.

Position angle and ellipticity profiles as a function of major-axis radius 
$r$ were derived from ellipse fits using the method of Bender \& M\"ollenhoff
(1987) and Bender et al.~(1988) as implemented in 
{\tt MIDAS\/} (Banse et al.~1988) by Bender and by Saglia (2003, private
communication).  Position angles are measured east of north.
Some profiles were extended using ellipse fits made by {\tt GASP\/} 
(Cawson 1983; Davis et al.~1985), which is slightly more robust at
low $S/N$ or when isophotes are incomplete at the edge of the field of
view.

Surface brightness cuts along the major axis, the minor axis,
and (for NGC 4593) the bar were produced using a program that averages 
pixel values in a 25$^\circ$-wide, pie-shaped wedge.  Therefore, more pixels
are included at large radii where the $S/N$ is low.  Masked pixels are left 
out of the average.  The cuts on opposite sides of the galaxy center were 
averaged. 

The various cut profiles were shifted in mag arcsec$^{-2}$ to match
up as well as possible.  The $K_{\rm s}$-band zeropoint for NGC 7690 was
derived from 2MASS, 5$^{\prime\prime}$- and 7$^{\prime\prime}$-radius,
circular-aperture photometry applied to the {\it HST\/} NICMOS F160W
($H$-band) image. The $K_{\rm s}$-band zeropoint for NGC 4593 was transferred
from 2MASS by measuring the galaxy magnitude in the Large Galaxy Atlas 
image (Jarrett et al.~2003) and in our AAT image within the largest 
(radius = 95$^{\prime\prime}$) aperture that fits within our image.
For NGC 7690, the $V$-band zeropoint was derived from aperture photometry 
by Wegner (1979).  For NGC 4593, the WFPC2 image is saturated, making 
calibration via aperture photometry problematic.  Instead we used the 
transformation from F606W to Johnson V given by Holtzman et al.~(1995), 
assuming that $V - I = 1.25$ from the aperture photometry by McAlary 
et al.~(1983) as tabulated by Prugniel \& H\'eraudeau (1998). 

The cut profiles were smoothed by averaging values in bins of width 0.04
in $\log {r}$.   Profiles were truncated where they were affected by seeing 
at small radii and low $S/N$ at large radii.

The $V - K_{\rm s}$ cut profile was created from calibrated $V$ and 
$K_{\rm s}$
composite profiles. The $V$ composite profile was created from a cut from
the Planetary Camera image inside a radius of 20$^{\prime\prime}$, and the 
cut from the Wide Field Camera beyond that.  Similarly, the $K_{\rm s}$ 
composite profile was created from the NICMOS profile inside 
10$^{\prime\prime}$ and the AAT profile outside that.

The $V - K_{\rm s}$ images were created by taking
the ratio of the sky-subtracted WFPC2 F606W $V$ and AAT $K_{\rm s}$ frames,
after convolving the WFPC2 frames with gaussians to match the AAT PSFs.
Figures 1 and 3 shows the logarithms of the ratio images.


\begin{figure*}[t]
\centerline{\psfig{file=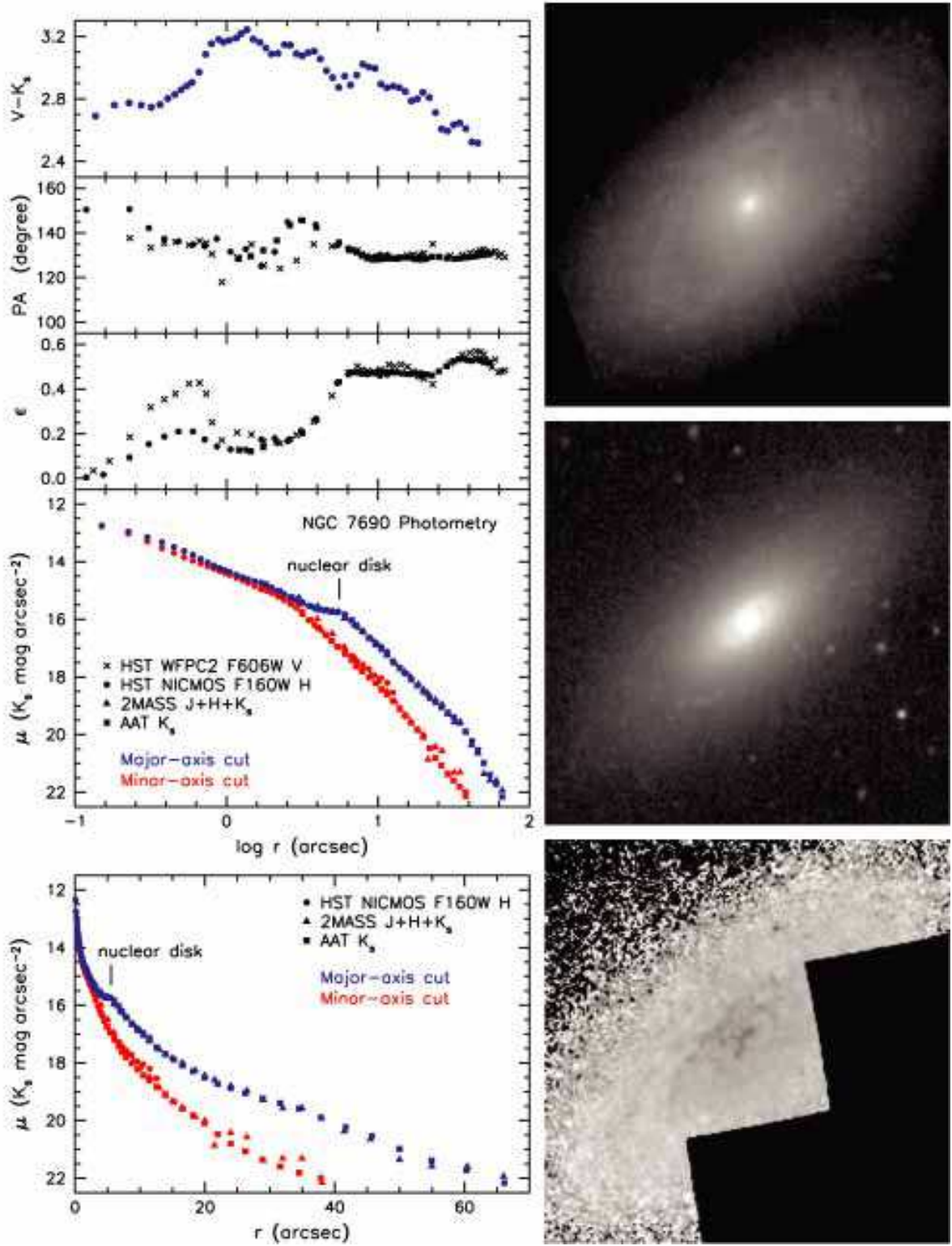,width=15.8cm,angle=0}}
\figcaption[n7690.ps]{
NGC 7690 pseudobulge -- top image: 18$^{\prime\prime}$ 
$\times$ 18$^{\prime\prime}$ {\it HST\/} NICMOS F160W image from 
Carollo et al.~(2002); middle: 91$^{\prime\prime}$ $\times$
91$^{\prime\prime}$ AAT $K_{\rm s}$-band image discussed in \S\ts2;
the images are shown at different logarithmic intensity stretches.
Bottom: $V - K_{\rm s}$ color image constructed from the above 
$K_{\rm s}$ frame and from a $V$-band (F606W) image taken with 
{\it HST\/} by Carollo et al.~(1998; they also show an
unsharp-masked image).  The intensity stretch is linear in mag 
arcsec$^{-2}$; white corresponds to $V - K_{\rm s} = 0.8$, and black 
corresponds to $V - K_{\rm s} = 3.5$.  The middle and bottom panels show
the same field of view.  North is up and east is at left in all images.
A large-scale \hbox{(9$^{\prime\prime}$ $\times$ 9$^{\prime\prime}$)}
$V - H$ color image obtained with {\it HST\/} is illustrated in Carollo 
et al.~(2002).  The left panels above show surface photometry as a function
of log radius (top) and linear radius (bottom).  Brightness and 
$V - K_{\rm s}$ profiles are 25$^\circ$-wide cuts along the major axis 
(PA = 129$^\circ$) and minor axis (PA = 39$^\circ$) color-coded as 
indicated in the key and with opposite sides of the galaxy averaged.  
Isophote ellipticities $\epsilon$ and major-axis position angles PA are 
derived from isophote fits to the images indicated in the key.  \label{}}
\end{figure*}


\section{NGC 7690}

    NGC 7690 is an unbarred, Sab galaxy illustrated in the Carnegie Atlas
(Sandage \& Bedke 1994).  In Tully (1988) group 61 $-0$ $+16$; its group's
mean recession velocity of $1495 \pm 59$ km s$^{-1}$ together with a Hubble
constant of 71 km s$^{-1}$ Mpc$^{-1}$ (Spergel et al.~2003) imply a 
distance of 21 Mpc. For a Galactic absorption of $A_B \simeq 0.045$
(Schlegel et al.~1998) and a total $B$-band apparent magnitude
of $B_{\rm T} = 13.0$ (de Vaucouleurs et al.~1991: RC3), the absolute 
magnitude is $M_B = -18.7$.  NGC 7690 is therefore a reasonably normal,
low-luminosity disk galaxy.  Its relatively early Hubble type implies that it
should have a substantial bulge (e.{\ts}g., Simien \& de Vaucouleurs 1986). 
In particular, NGC 7690 is earlier
in Hubble type than the relatively sharp transition observed to occur
at Sb/Sbc between early-type galaxies that mostly contain classical 
bulges and late-type galaxies that mostly contain pseudobulges (see
KK for a review).  NGC 7690 nevertheless proves to contain a well 
developed pseudobulge.

      The pseudobulge of NGC 7690 was discovered by Carollo et al.~(1998); 
their $V$-band image is included in Figure~13 of KK.
An {\it HST\/} F160W image from Carollo et al.~(2002) is 
shown in the upper-right panel of our Figure 1.  Qualitatively, the center 
of NGC 7690 looks disk-like, so much so that in the
``Bulge?''~column of Table 2 in Carollo et al.~(1998), they do not 
even list NGC 7690 as having an ``IB'' = ``irregular bulge'', which is 
essentially equivalent to what we call a pseudobulge.  Instead, the entry
is ``No?''.  NGC 7690 was chosen for this paper because we wanted to 
investigate the (pseudo)bulge quantitatively with near-infrared 
surface photometry.

      The results are shown in Figure 1.  The blue and red $\mu(r)$ 
points are major- and minor-axis brightness cuts calculated as discussed 
in \S\ts2.3.  Excellent agreement between the AAT and 2MASS profiles 
provides an important check of our reduction procedures.  The
$\epsilon(r)$ and PA$(r)$ points are isophote ellipticity and position 
angle profiles calculated by fitting elliptical isophotes to the images 
identified in the key.  Calculations of the $V - K_{\rm s}$ color profile 
(top-left plot) and image (lower-right) are discussed in \S\ts2.3.  The 
profiles are plotted against $\log {r}$ to illustrate the behavior at 
small and large radii and also against linear radius $r$ so that an
exponential outer disk -- and departures of the inner profile above it -- 
can easily be recognized.

      Fig.~1 shows three quantitative signatures of a pseudobulge:

      First, the central component illustrated by Carollo and collaborators
forms a clear shelf in the brightness distribution at major- and minor-axis
radii of 6$^{\prime\prime}$ and 3$^{\prime\prime}$, respectively.  That is, it
has the morphology of a lens (Freeman 1975; Kormendy 1979, 1981; Buta \& Combes
1996), not a bar.  A bar would form a shelf in the major-axis profile only.  
In contrast, a lens\footnote{Lens components should not be confused with
lenticular galaxies, i.{\ts}e., with the name for S0 galaxies (Sandage 1961)
that is adopted by de Vaucouleurs and collaborators (e.{\ts}g., RC3).  Many 
S0 galaxies do not contain lenses, and many Sa and later-type galaxies do 
contain lenses.  This confusion is unfortunate, but it is thoroughly embedded
in the literature.} is defined as a shelf of nearly constant surface brightness
seen along both the major and minor axes.  Lenses have intrinsic axial ratios
$\sim 0.9 \pm 0.05$ in the equatorial plane (see the above papers), whereas
bars in early-type galaxies have axial ratios of $\sim 0.1$ to 0.3 in the 
equatorial plane (see Sellwood \& Wilkinson 1993 for a review).  In other
words, the shape distributions of lenses and bars do not overlap.  
Lenses and bars are physically different.  In fact, barred galaxies often 
contain both components, with the bar filling the lens in one dimension
(see the above papers).  However, unbarred galaxies can also have strong
lenses (e.{\ts}g., NGC 1553: Freeman 1975; Kormendy 1984).  Note further that 
lenses and ``inner rings'' (bright rings that generally encircle the end
of the bar) are also different (e.{\ts}g., Sellwood \& Wilkinson 1993;
Buta \& Combes 1996): inner rings are relatively dark inside.  Also, rings
are narrow in the radial direction and therefore must have small radial
velocity dispersions, whereas lenses are observed to have large radial velocity 
dispersions (e.{\ts}g., Kormendy 1984).  The main bars in early-type galactic 
disks have radii of $\sim$ 1.4 exponential scale lengths (Erwin 2005), and
-- as noted above -- the radii of lenses and inner rings are closely similar to 
the sizes of their associated bars.  Distinctly smaller versions of all of these
phenomena -- nuclear bars, nuclear lenses (often but not always associated
with nuclear bars), and nuclear rings -- also occur, usually but not always 
in galaxies that also contain larger, ``main'' bars or ovals.  We belabor 
these points because lenses are less well known in the astronomical
community than bars or inner rings.  The 6$^{\prime\prime}$ $\times$ 
3$^{\prime\prime}$ shelf in the brightness distribution of NGC 7690 is 
a nuclear lens without an associated nuclear or main bar.  

      The nuclear lens is 
1.5 to 2 $K_{\rm s}$ mag arcsec$^{-2}$ higher in surface brightness than the 
inward extrapolation of the outer disk's almost-exponential brightness 
profile.  It therefore forms part of what would conventionally be 
identified as the bulge.  We would like to classify a bulge as the E-like
part of a galaxy.  In practice, this is not straightforward for
non-edge-on galaxies.  Therefore, Carollo et al.~(1999) adopt the 
surrogate definition that a bulge is the central part of 
the galaxy that is brighter than the inward extrapolation of the outer
disk's exponential brightness profile.  By this definition, the nuclear 
lens in NGC 7690 is part of the galaxy's bulge.  But bulges and elliptical
galaxies generally have simple brightness profiles consisting of a single
S\'ersic (1968) $\log {I} \propto r^{1/n}$ profile (possibly with a cuspy 
core) and not a shelf in the brightness profile such as the one in NGC 7690. 

      The second pseudobulge indicator is that the nuclear lens -- the shelf 
in the inner brightness profile --
is clearly a disk. This was already evident from the images in Carollo et
al.~(1998, 2002), as shown here in Figure 1.  The nuclear disk in the 
top-right panel has the same apparent flattening and position angle as the 
outer disk shown in the middle image panel.  We see this result 
quantitatively in the $\epsilon$ and PA plots.  The $r = 6^{\prime\prime}$ 
shelf in the brightness profile has precisely the same ellipticity and PA 
as the disk farther out.  It is not plausible that this is due to internal
absorption, because the $V$-band ellipticities and position angles (crosses
in Figure 1) agree with the $K_{\rm s}$-band ones at all radii that are 
relevant to our conclusions.  Since the $K_{\rm s}$-band extinction is 
about one-tenth as big as the $V$-band extinction (e.{\ts}g., Cox 2000), 
this implies that extinction does not significantly affect the 
parameters at either wavelength, unless unrealistically gray extinction is
postulated.  From $r \simeq 6^{\prime\prime}$ outward, the ellipticity and
PA change very little, consistent with a nearly  circular main disk at 
nearly constant inclination.  We conclude that the nuclear lens, i.{\ts}e.,
the outer part of what would conventionally be identified as the bulge of 
NGC 7690, is as flat as the galaxy's disk.  This is the most clearcut 
signature of a pseudobulge in NGC 7690 (see KK for pseudobulge 
classification criteria).

      The surface brightness of the nuclear disk is nearly constant
interior to its sharp outer edge.  Not surprisingly, the ellipticity 
drops rapidly inward as the profile of this highly-inclined galaxy 
becomes dominated by a less flattened center.  This feature may be but 
is not necessarily a classical bulge.  It would be no surprise if the
pseudobulge part of NGC 7690 were embedded in a classical bulge,
because secular evolution can build a pseudobulge inside a pre-existing
bulge (see KK).  But several arguments suggest that a classical bulge 
is not the main feature near the center.  First, the PA profile shows 
a clear twist centered at $r \simeq 3^{\prime\prime}$.  The ellipticity 
starts to drop suddenly toward the center at the radius $r \simeq 
6^{\prime\prime}$ where the twist starts, and it continues to drop
throughout the radius range of the twist. This combination is a characteristic
of a weak bar or weak spiral arms. Bars and spiral arms are disk features.
Thus, even interior to its sharp outer edge, the signs are that the 
galaxy is dominated by a disk.

      The third pseudobulge indicator is the lack of a steep color
gradient between a bluish disk and a red and much older bulge.  The
$V - K_{\rm s}$ color image (bottom-right in Figure 1) and profile both 
show that, apart from a few irregular dust features that are easily masked
out in the photometry, there is no significant color difference between 
the inner part of the outer disk and the nuclear disk.  Indeed, the 
$V - K_{\rm s}$ profile is completely continuous from the disk-dominated 
outer galaxy to the bulge-dominated inner galaxy.  This is an example of
a general phenomenon: (pseudo)bulge and disk colors correlate
(Peletier \& Balcells 1996; Gadotti \& dos Anjos 2001).  Bulges
and disks both show large ranges in colors, but  ``bulges are more like 
their disk[s] than they are like each other'' (Wyse et al.~1997).
Courteau et al.~(1996) and Courteau (1996) interpret such 
correlations as products of secular dynamical evolution.  As noted above,
there is no sign in NGC 7690 of a discontinuity in stellar population 
between an old, non-star-forming bulge and a younger disk, as one would
expect for a classical bulge (Sandage \& Bedke 1994).

      Note that the small color gradient from outer disk to nuclear 
lens implies, since $K_{\rm s}$ extinctions are so much smaller than 
$V$ extinctions, that the shelf in the brightness profile is a real
feature in the stellar density and not an artifact of absorption.

      Several classification criteria agree, so it seems safe to conclude 
that NGC 7690 contains a pseudobulge.  This may or may not coexist with 
a small classical bulge -- we cannot be certain from the data at hand,
although no observation points compellingly to a classical bulge component.

      We conclude that NGC 7690 is an example of secular evolution in 
action.  This is particularly interesting because the
Sab galaxy NGC 7690 is earlier in type than the galaxies\footnote{We
note, however, that even S0s can contain pseudobulges (KK).}
that most commonly contain pseudobulges. Also, there is no obvious 
engine for secular evolution, such as a bar, oval distortion, or global
spiral structure.  NGC 7690 therefore contributes to the accumulating
evidence that secular evolution occurs in a wide variety of galaxies.



\begin{figure*}[t]
\centerline{\psfig{file=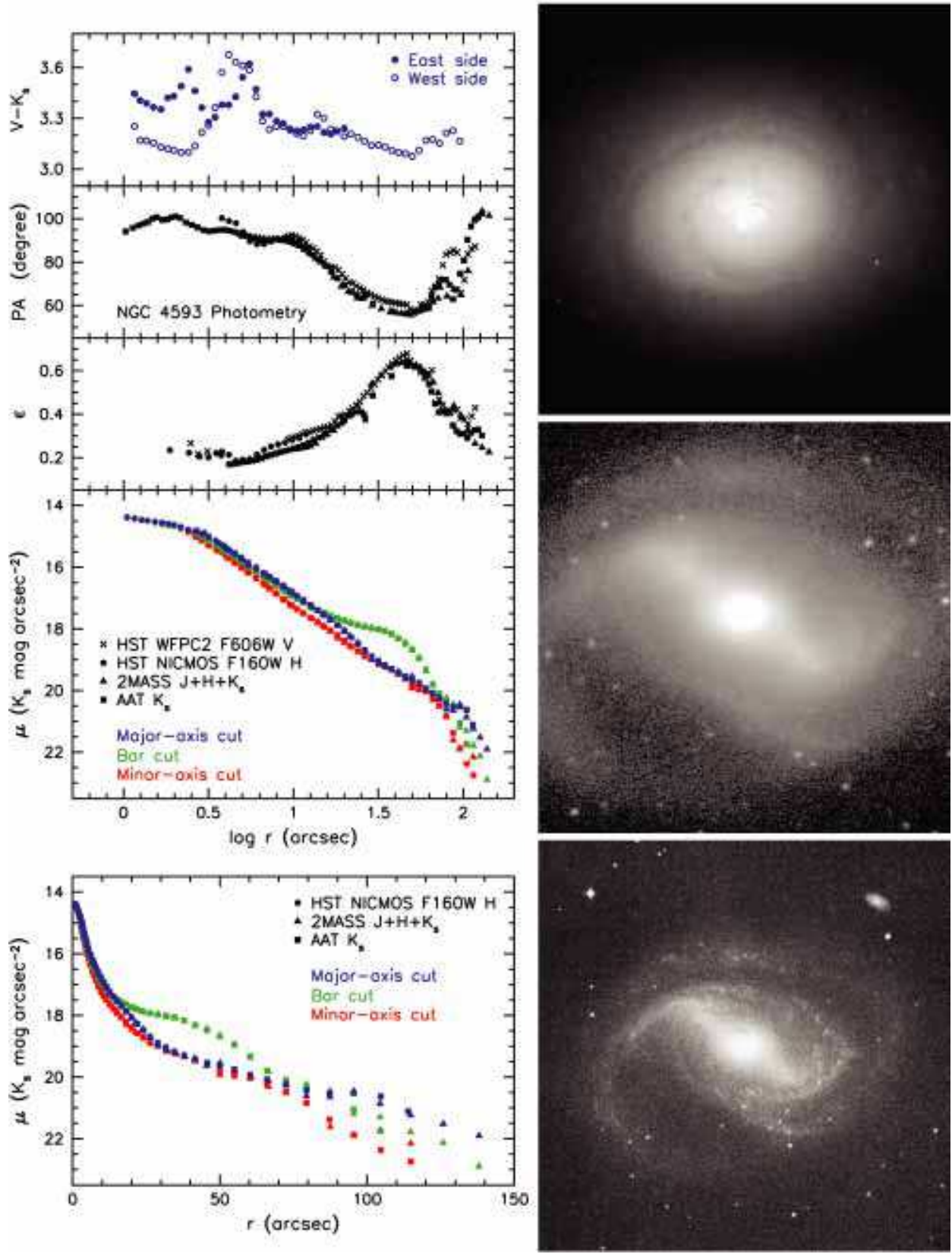,width=16.5cm,angle=0}}
\figcaption[]{
NGC 4593 pseudobulge -- top image: 18$^{\prime\prime}$ 
$\times$ 18$^{\prime\prime}$ {\it HST\/} NICMOS F160W image from 
Carollo et al.~(2002); middle: 191$^{\prime\prime}$ $\times$
191$^{\prime\prime}$ AAT $K_{\rm s}$-band image discussed in \S\ts2;
the images are shown at different logarithmic intensity stretches.
Bottom: 284$^{\prime\prime}$ $\times$ 284$^{\prime\prime}$ $B$-band
image from the Carnegie Atlas of Galaxies (Sandage \& Bedke 1994).
North is up and east is at left in all images.  The left panels show 
surface photometry as a function of log radius (top) and linear radius
(bottom).  Brightness profiles are 25$^\circ$-wide cuts along the major
axis (PA = 101$^{\circ}$), bar (PA = 57$^{\circ}$), and minor axis 
(PA = 11$^{\circ}$) with opposite sides of the galaxy averaged.  Isophote
ellipticites $\epsilon$ and major-axis position angles PA are derived from 
isophote fits to the images indicated in the key.
\label{}}
\end{figure*}


\section{NGC 4593}

     NGC 4593 is a structurally normal SBb galaxy (Sandage 1961; Sandage 
\& Bedke 1994).  We adopt a distance of 35 Mpc (Tully 1988 group 
11 $-29$), then $B_{\rm T} = 11.67$ (RC3) implies that $M_B = -21.2$.  
NGC 4593 is therefore a high-luminosity disk galaxy.  Moreover, it has a 
well developed engine for secular evolution in the form of a high-amplitude 
bar.  Secular evolution is likely to be rapid.  Our photometry of NGC 4593 
proves to be consistent with this expectation (Figure 2).

\subsection{Near-Infrared Photometry: Evidence for a Pseudobulge}

      NGC 4593 is a well known Seyfert I galaxy (Lewis et al.~1978; 
MacAlpine et al.~1979).  The nuclear point source is 
very bright ({\it HST\/} F606W mag $\sim 17.1$ at $r \leq 0\farcs09$); it
completely swamps the central brightness distribution of the galaxy
(MRK 1330 in Figure 1 of Malkan et al.~1998).
Since this is not star light, it is important that we subtract it in our 
analysis of the brightness distribution.  We did this for the {\it HST\/}
NICMOS image by subtracting a PSF calculated with {\tt Tiny Tim\/}
(Krist \& Hook 2004) and scaled in total intensity to remove the diffraction 
spikes and other small-scale PSF structure in the image as well as possible. 
The result is the image in the upper-right panel of Figure 2; it was used to
calculate the NICMOS points in the profile plot panels.  Of course, it is not
possible to recover the stellar brightness distribution very close to the 
center.  We were conservative and truncated the brightness profile at
$r$ \lapprox \ts1$^{\prime\prime}$. Uncertainties in the {\it HST\/} PSF subtraction
do not affect our conclusions.  It is important to note that this would not be
true at ground-based resolution.  The AAT $K_{\rm s}$ profile does not show
the profile kink at $\log {r} = 0.5$ (see Figure 2 and discussion, below), 
because the central, fainter part of the profile is filled in by the 
seeing-convolved Seyfert nucleus.

      A comparison of the major-axis, bar, and minor-axis brightness
cuts then shows that the ``bulge'' dominates the brightness distribution
at major-axis radii $r < 20^{\prime\prime}$.  Is this a classical
bulge or is it a pseudobulge?

      The ellipticity profile shows that NGC 4593 contains a
pseudobulge, with no sign of a classical bulge component.  The
apparent ellipticity of the outer disk is $\epsilon \simeq 0.25 \pm 0.05$.
Remarkably, the ellipticity of the bulge is also $0.25 \pm 0.05$
over the whole radius in which we can measure it; i.{\ts}e.,
interior to the bar and exterior to the region clobbered by the 
Seyfert nucleus.  Note again that {\it HST\/} resolution is important: 
the extra smoothing caused by the ground-based PSF makes the isophotes 
as measured with the AAT significantly rounder than those measured with 
{\it HST\/} NICMOS.  But the NICMOS data show that the pseudobulge of 
NGC 4593 is as flat as its disk.  

      In the absence of PA information, it might be possible that the
``bulge'' of NGC 4593 is really a nuclear bar.  Nuclear bars are common
in barred galaxies, and since they have arbitrary 
position angles with respect to their associated main bars, an
elongated ``bulge'' could be a vertically thick nuclear bar rather
than a vertically thin disk.  However, the PA is essentially the same
at small radii as in the outer disk.  For this to be caused by a nuclear
bar would require the added accident that the nuclear bar is aligned 
with the major axis of the galaxy.  This is not impossible, but the
more likely explanation is that the bulge is nearly circular in its
equatorial plane and as flat as the disk.

      In either case, the observations imply a pseudobulge.  The cleanest
signature of a well developed pseudobulge is that it is very flat.  However, 
since bars are disk phenomena, the observation of a nuclear bar would also be
evidence for a pseudobulge.  Both classification criteria are reviewed in KK.

      A second feature of the profiles in Figure 2 is consistent~with a 
pseudobulge but does not by itself prove that one is present. 
The (pseudo)bulge profile has a kink at $\log {r} \simeq 0.5$ 
($r \simeq 3^{\prime\prime}$).  Such features are not seen in 
classical bulges or ellipticals.  The above
radius is too large for the kink to be a cuspy core like those
in ellipticals (see Lauer et al.~1995; Faber et al.~1997).
Pseudobulges are a consequence of more complicated physics
than the violent relaxation and dissipation that builds ellipticals.
If they grow by star formation in gas that has been transported
to the center, exactly how the star formation proceeds and what kind of 
density profile it produces are controlled by a complicated interplay 
between star formation (as described by a Schmidt 1959 -- 
Kennicutt~1998a,{\ts}b law) and the factors -- e.{\ts}g., resonances -- that 
determine where the gas stalls.  Nuclear star formation rings seen in many 
barred and oval galaxies are a hint that the 
stellar density that ``rains out'' of the gas distribution may have more 
complicated radial features than do elliptical galaxies, in which violent 
relaxation smooths the density in radius.  So pseudobulges are likely 
to have a larger variety of profile shapes than do ellipticals.
We see two examples of this variety in the present paper.


\begin{figure*}[t]
\centerline{\psfig{file=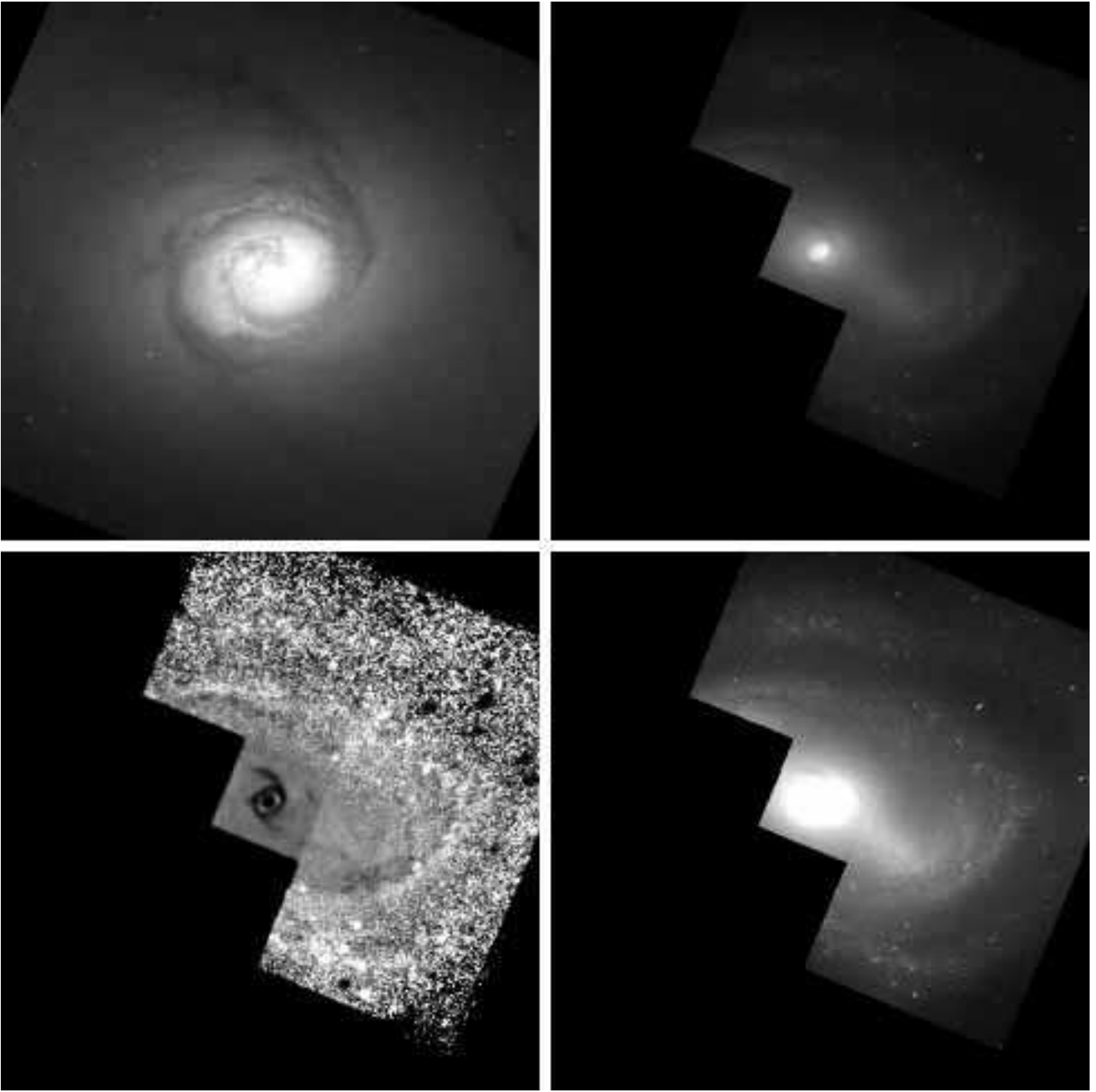,width=18.1cm,angle=0}}
\figcaption[]{
NGC 4593 $V$-band {\it HST\/} images (top left, top right, and bottom
right) from Malkan et al.~(1998) and $V - K_{\rm s}$ color image 
(bottom left).  The PSF of the Seyfert nucleus was modeled using 
{\tt Tiny Tim\/} and has been subtracted. 
The $V$-band images are shown at different logarithmic intensity
 stretches.  The intensity stretch of the $V - K_{\rm s}$ color image is 
linear in mag arcsec$^{-2}$; white corresponds to $V - K_{\rm s} = 2.9$, 
and black corresponds to $V - K_{\rm s} = 4.7$. 
The top-left panel is $30^{\prime\prime} \times 30^{\prime\prime}$ and
can be compared directly with the $18^{\prime\prime} \times 18^{\prime\prime}$
top-right panel in Figure 2.  Both 
show a dust feature starting tangential to the well defined dust ring and 
spiraling in to the Seyfert nucleus at the center.  The other three images 
are 191$^{\prime\prime}$ $\times$ 191$^{\prime\prime}$ and show the same
field of view as the middle image panel in Figure 2.  They illustrate 
the relationship between the dust ring with the bar.  Almost radial 
dust lanes in the bar turn into spirals close to the center and become
tangential to the dust ring.  All of the dust lanes are very red, as 
shown by the $V - K_{\rm s}$ image and by the color cuts in the top-left 
panel of Figure 2.  
\label{}}
\end{figure*}


\subsection{Comparison of $V$ and $K_{\rm s}$ Images: The Inner Dust Ring
            as Evidence for Episodic Pseudobulge Growth}

      Dust is more important in NGC 4593 than in NGC 7690.  Therefore 
we verify in this subsection that dust absorption does not 
compromise the above conclusions.  Examination of the dust features
also leads to a new result, namely a hint that the starbursts
that contribute most to pseudobulge growth may be episodic.

      An inner dust spiral is faintly visible in the top-right panel
of Figure 2.  Its relationship with the rest of the galaxy is made clear
in Figure 3, which shows $V$-band images that can be compared with 
$K_{\rm s}$ images in Figure 2.  Figure 3 also shows a $V - K_{\rm s}$
color image; major-axis cuts through this image are shown in the
$V - K_{\rm s}$ color profile in the top-left panel of Figure~2.
We show separately the color profiles east and west of the center;
the red ``peaks'' identify where narrow dust features cross the major
axis.

      The $V$-band {\it HST\/} PC image (Figure 3) shows a strong dust 
ring with major-axis radius $\simeq$ 5$^{\prime\prime}$.  It causes the 
red ring that is the most obvious feature of the $V - K_{\rm s}$ color
image.  It is also seen as the red peak in both the
east and west color profiles at $\log {r} \simeq 0.7$ in Figure 2.  
Interior to the dust ring, the faint dust spiral seen in Figure 2 is,
not surprisingly, much more obvious at $V$ band in Figure 3.  Its
crossings of the major axis are seen in the color profile in Figure 2
as $V - K_{\rm s}$ maxima at $\log {r} \simeq 0.6$ on the west side 
(open circles) and at $\log {r} \simeq 0.4$ on the east side (filled 
circles).  Fortunately, the dust spiral proves to be very narrow.  Over 
a substantial radius range interior to the dust ring, the colors of the
pseudobulge on the west side of the center are closely similar to those
in the outer disk.  There is no sign of significant reddening or (by
inference) absorption at these radii.  The minor axis on the south side 
of the center is similarly free of absorption except in the dust spiral.
We used one-sided brightness cuts in these two directions to measure 
$V$-band ellipticities at $2\farcs5 \leq r \leq 4^{\prime\prime}$.
These are shown as the crosses in the $\epsilon$ profile in Figure 2.
The $V$-band ellipticities, carefully measured to avoid the
dust, agree with the $K_{\rm s}$ values.  The same is true at radii
outside the dust ring, where dust absorption is small enough so that
$V$-band isophote fits are possible and both $\epsilon$ and PA can be
measured.  As in the case of NGC 7690, we conclude that the 
near-infrared measurements, which are much less affected by internal 
absorption, are reliable.  This confirms our conclusions about the 
flattening of the pseudobulge.

       We emphasize again that, except in the very red but narrow 
dust ring and in the central arcsec, the pseudobulge in NGC 4593 is the
same color, $V - K_{\rm s} \simeq 3.2 \pm 0.1$, as the outer disk at 
$r \simeq 100^{\prime\prime}$.  Both are about 0.4 mag redder in 
$V - K_{\rm s}$ than is NGC 7690.  These results again are consistent
with Wyse et al.~(1997): Bulges and disks both show large ranges in colors,
but ``bulges are more like their disk[s] than they are like each other.''
The red color of NGC 4593 has been noted previously (Santos-Lle\'o 
et al.~1995; Shaw et al.~1995).  The inner 
ring and the spiral arms between $r \simeq 30^{\prime\prime}$ and
80$^{\prime\prime}$ are slightly bluer than the rest of the disk.  Inner 
ring formation is part of the canonical secular evolution picture, and 
spiral structure is a signature of outward angular momentum transport. 
But the dust-free parts of the pseudobulge interior to the dust ring
are as blue as the outer sprial structure. So the color data are 
consistent with slow pseudobulge growth.

      That growth may be episodic.  The $V - K_{\rm s}$ color image in
Figure 3 is instructive.  It and the $V$-band image both show a dust lane 
on the rotationally leading side of the bar that becomes curved and 
eventually -- near the center -- tangent to the nuclear dust ring.  
Simulations of gas response to a barred potential
(see especially Athanassoula 1992; KK provide a review) suggest that 
such dust lanes occur at shocks where the gas and dust are compressed.
Velocity discontinuities across dust lanes observed in H{\ts}I
(e.{\ts}g., Lindblad et al.~1996; Regan et al.~1997) and H{\ts}II 
(Zurita et al.~2004)
provide compelling support.  If shocks are present, it is
inevitable that gas loses energy and falls toward the center.  All this
is central to the secular evolution picture.  Intense starbursts, often
in the form of nuclear rings, are commonly associated with the above
phenomena (KK provide a detailed review) and are widely interpreted as the 
result of the gas inflow.  This star formation is part of pseudobulge 
growth.  Interestingly, in NGC 4593, we see a nuclear dust ring, not
a starburst ring. This suggests that gas is accumulating as a result
of bar-driven inflow but that it is not currently starbursting (see
also Oliva et al.~1999).  Since high gas densities favor high star 
formation rates (Schmidt 1959; Kennicutt 1998a, b), it is reasonably 
to assume that the dust and associated gas ring will turn into a 
star-forming ring some time in the future.

      Finally, we return to the one-armed, spiral dust lane that continues 
inward from the dust ring to the central region that is dominated by the 
Seyfert nucleus (Malkan et al.~1998; Figure 3 here).  Spiral dust lanes
interior to star formation rings are also seen in many galaxies (see KK).  
Elmegreen et al.~(1998) suggest that they are a sign that some gas continues 
to sink toward the center even interior to the radius at which most gas 
stalls.  Additional pseudobulge growth and feeding of the Seyfert nucleus 
are plausible consequences.

      In summary, the $V$-band {\it HST\/} image and the $V - K_{\rm s}$
color image in Fig.~3 further support our conclusion that bar-driven 
gas inflow continues to grow a pseudobulge in NGC 4593.  They also 
emphasize a new aspect to the secular evolution picture: the strong 
starbursts that are seen in many galaxies and that are interpreted as 
an important part of pseudobulge growth may, at least in some galaxies, 
be episodic.

\section{Possible Effects of Galaxy Interactions and Mergers}

      Kannappan et al.~(2004) suggest that accretions of
small galaxies are an alternative way to build cold, disky subsystems
in galaxies.  They find a correlation between blue-centered, star-forming 
bulges and evidence of tidal encounters with galaxy neighbors.  Such
processes must happen; embedded counterrotating components provide
the clearest examples (e.{\ts}g., NGC 4826, Braun et al.~1994; Rubin 1994; 
Walterbos et al.~1994; Rix et al.~1995; Garc\'\i a-Burillo et al.~2003). 
The observation that so many prominent pseudobulges occur in barred and 
oval galaxies -- that is, ones that contain obvious engines for secular
evolution -- is one argument that galaxy interactions do not 
produce most pseudobulges (KK discuss others).  These arguments are 
statistical; they do not much constrain individual galaxies.  It is 
reasonable to ask (as the referee did): could accreted material 
account for our observations?

      The answer for NGC 7690 is ``maybe, but there is no evidence for
this''.  The answer for NGC 4593 is ``probably no''.

      First, we need to understand what kind of accretion is possible.
Major mergers that happened recently in the history of both galaxies are,
we believe, excluded.  T\'oth \& Ostriker (1992) emphasized that disks 
are fragile; they are easily destroyed by dynamical stirring produced
by even a low-mass projectile.  Both galaxies have flat components near
their centers.  Accretion of a high-mass galaxy or one that already has a
bulge would destroy such a subsystem and leave behind a big classical
bulge that we do not see.  This is especially relevant in NGC 4593,
which remains flat all the way in to the Seyfert nucleus.  What could most 
safely be accreted is a bulgeless, gas-rich dwarf galaxy; its fluffy 
stellar distribution would get torn apart by tidal effects at large radii, 
and its gas could dissipate its way into the central regions. 

\subsection{NGC 7690}

     The galaxy is isolated on the Digital Sky Survey.  The closest 
galaxies listed by the NASA/IPAC Extragalactic Database (NED) that have 
measured recession velocities within 1000 km s$^{-1}$
of that of NGC 7690 are ESO 240-G12
(projected distance = 54\farcm5 = 49 radii of NGC 7690) and ESO 240-G4
(projected distance = 59\farcm0 = 53 radii of NGC 7690), where the 
radius of NGC 7690 at 25 $B$ mag arcsec$^{-2}$ is 1\farcm1 (RC3).  In an 
H{\ts}I survey to look for galaxy pairs (Chengalur \etal 1993), it was 
not listed as a pair.  It has a normal, two-horned, single-dish H{\ts}I
velocity profile with almost no asymmetry (Davies et al.~1989; Chengalur 
et al.~1993); many isolated, late-type galaxies have more asymmetric 
velocity profiles.  No asymmetry is seen in {\it The Carnegie Atlas of
Galaxies\/} (Sandage \& Bedke 1994), and we see none in our deep AAT
images.  No interaction appears to be in progress.  We cannot exclude
a past minor accretion event, but no smoking gun points to one.

\subsection{NGC 4593}

     There are clear signs that NGC 4593 is interacting with PGC 42399.
Any such interaction is fast -- the velocity difference of 320 km s$^{-1}$
(NED) is large compared with plausible galaxy rotation velocities.  This 
does not favor a strong interaction.  Also, NGC 4593 is brighter than 
PGC 42399 by a factor of 7.7 (NED), so the perturber is not very massive.
On the other hand, its projected distance from NGC 4593 is only about two 
disk radii.  The spiral structure of NGC 4593 is slightly distorted toward 
PGC 42339.  Some tidal stretching and possibly some tidal tickling of the 
wave patterns (spiral arms and bar) in the galaxy are plausible.

     Still, the structure of the galaxy is -- apart from the above --
completely normal (Sandage \& Bedke 1994).  The galaxy has an inner
ring at the end of the bar, as do many other barred galaxies.  Such
rings are by now reasonably well understood as products of long-term,
bar-driven secular evolution.  Inner rings are signs that evolution has 
been going on for a long time.  Simulations suggest that the SB(r) phase 
comes after the SB(s) phase, after outward angular momentum transport has 
slowed the pattern speed of the bar and after the bar has had time to 
rearrange disk gas into an inner ring (see KK for a review).
This is relevant because simulations also suggest that tidal interactions
can trigger bar formation (e.{\ts}g., Noguchi 1987, 1988; Gerin et al.~1990; 
Barnes \& Hernquist 1991; Elmegreen et al.~1991, 
although see also Sellwood 2000).  Therefore, although it is tempting
to wonder whether the present interaction has something to do with the
barred structure of NGC 4593, the mature morphology of the SB(r) 
structure suggests that the galaxy has been barred and evolving secularly 
for a long time.  The structure of the dust features -- a nearly radial 
dust lane on the leading side of the bar becoming tangent to a nuclear 
dust ring -- are a clear and clean sign of evolution in action
(Athanassoula 1992).  

       Finally, the spiral structure of NGC 4593 is not 
unusual or suggestive of the influence of the interaction.  One arm 
starts on the inner ring at the end of the bar and the other starts 
``about 15 degrees downstream'' from the end of the bar (Sandage \& Bedke 
1994).  This behavior is similar to that of NGC 2523, another 
prototypical SB(r) galaxy (Sandage 1961).  NGC 2523 is not asymmetric, and 
it is more isolated than NGC 4594 (the nearest substantial companion,
NGC 2523B, is more than 7 NGC 2523 radii away in projection and has a 
recession velocity 365 km s$^{-1}$ larger than that of NGC 2523; RC3).  

      Therefore we believe that the present weak and temporary interaction
with PGC 42399 is not responsible for the main features of the structure 
of NGC 4593, including the pseudobulge.

\section{Conclusion}

      NGC 7690 (Sab) and NGC 4593 (SBb) provide clean examples of 
relatively early-type galaxies whose ``bulges'' are more disk-like
than any elliptical galaxy.  In particular, elliptical galaxies are 
never flatter than axial ratio $\simeq 0.4$ (Sandage et al.~1970;
Binney \& de Vaucouleurs 1981; Tremblay \& Merritt 1995), whereas 
part (NGC 7690) or essentially all (NGC 4593) of the bulges of the present 
galaxies are as flat as their outer disks.  We conclude that both galaxies 
contain pseudobulges -- that is, high-density, central components that were 
made out of disk gas by secular evolution.  In NGC 7690, blue colors imply
that star formation and hence pseudobulge growth are still in progress.
In NGC 4593, gas appears currently to be accumulating in a ring that
plausibly will form stars in the future.
Our results are examples of the general conclusion (see KK for a review)
that secular dynamical evolution occurs naturally and often in disk galaxies,
whether (NGC 4593) or not (NGC 7690) an engine for the evolution is 
readily recognized.

      To further investigate secular evolution, a desirable next step would 
be to quantify bar strengths by measuring~bar~torques, the ratio
of the bar-induced, tangential force to the mean radial force as a function 
of radius (Buta \& Block 2001; Laurikainen \& Salo 2002; Block et al.~2001, 
2004; Laurikainen et al.~2004).  This ratio can be as high 
as 0.6 in strong bars, emphasizing how efficient bars can be in 
redistributing angular momentum.  It would particularly be worthwhile to 
look for correlations between maximum bar torques and quantifiable 
consequences of secular evolution (e.{\ts}g., ring-to-disk and
pseudobulge-to-disk mass ratios) in a statistically representative sample 
of galaxies.

\acknowledgments

      JK is sincerely grateful to Mrs.~M.~Keeton and the Board of
Trustees of the Anglo American Chairman's Fund for the financial 
support that made possible his visit to South Africa during which 
this paper was written.  He also thanks the Cosmic Dust Laboratory and 
the School of Computational and Applied Mathematics of the University
of the Witwatersrand for their hospitality.  JHK acknowledges support
from the Leverhulme Trust in the form of a Leverhulme Research
Fellowship.  We thank the referee for a careful reading that led to 
substantial improvements to this paper.
JHK and EA wish to thank the staff of the AAT, and in particular
Dr.~Stuart Ryder, for their excellent support during his observing 
run with IRIS2. The authors are grateful to Ron Buta, to Bruce 
and Debra Elmegreen, and to Ivanio Puerari for making data from their
programs with DLB and JHK available for the present work.  We also
thank Roberto Saglia for his version of the Bender photometry
pipeline.
This research has made use of the NASA/IPAC Extragalactic Database 
(NED), which is operated by the Jet Propulsion Laboratory, California 
Institute of Technology, under contract with NASA. This research also
used the HyperLeda electronic database at 
{\tt http://www-obs.univ-lyon1.fr/hypercat} and the image display
tool {\tt SAOImage DS9} developed by Smithsonian Astrophysical Observatory.


\end{document}